\documentclass[useAMS,usegraphicx,usenatbib]{mn2e}

\newcommand{\oii}{\mbox{${\rm [OII]\lambda3727}$}}
\newcommand{\oiiis}{\mbox{${\rm [OIII]\lambda5007}$}}
\newcommand{\oiii}{\mbox{${\rm [OIII]\lambda4959,\lambda5007}$}}

\newcommand{\sii}{\mbox{${\rm [SII]\lambda6717,\lambda6731}$}}
\newcommand{\halpha}{\mbox{${\rm H\alpha}$}}
\newcommand{\hbeta}{\mbox{${\rm H\beta}$}}

\newcommand{\nii}{\mbox{${\rm [NII]\lambda6584}$}}

\newcommand{\slantfrac}[2]{#1\!\left/#2\right.}
\newcommand{\snr}{$\slantfrac{S}{N}$}

   \newcommand{\aap}{A\&A}
\newcommand{\aj}{AJ}         \newcommand{\apj}{ApJ}
\newcommand{\apjl}{ApJ}      \newcommand{\apjs}{ApJS}
\newcommand{\mnras}{MNRAS}   
     \newcommand{\pasp}{PASP}

\title{The environmental dependence of the chemical properties 
of star-forming galaxies}

\author[Mouhcine et al.]{M.~Mouhcine$^1$\thanks{Isaac Roberts 
Fellow}, I.~K.~Baldry$^1$, S.~P.~Bamford$^2$\\
$^{1}$Astrophysics Research Institute, Liverpool John Moores 
University, Twelve Quays House, Egerton Wharf, Birkenhead, 
CH41 1LD, UK.\\
$^2$ Institute of Cosmology and Gravitation, University of 
Portsmouth, Mercantile House, Hampshire Terrace, Portsmouth, 
PO1 2EG, UK.}

\date{Accepted ?. Received ?; in original form ?}

\pagerange{\pageref{firstpage}--\pageref{lastpage}} 
\pubyear{2007}

\begin{document}

\maketitle

\label{firstpage}

\begin{abstract}

We use a $0.040 < z < 0.085$ sample of $37\,866$ 
star-forming galaxies from the Fourth Data Release of the 
Sloan Digital Sky Survey to investigate the dependence of 
gas-phase chemical properties on stellar mass and environment.  
The local density, determined from the projected distances to 
the fourth and fifth nearest neighbours, is used as an 
environment indicator. Considering environments ranging from 
voids, i.e., $\log \Sigma\la -0.8$, to the periphery of galaxy 
clusters, i.e., $\log\Sigma \approx 0.8$, we find no dependence 
of the relationship between galaxy stellar mass and gas-phase 
oxygen abundance, along with its associated scatter, on local 
galaxy density. However, the star-forming gas in galaxies shows 
a marginal increase in the chemical enrichment level at a fixed 
stellar mass in denser environments. Compared with galaxies 
of similar stellar mass in low density environments, they are 
enhanced by a few per cent for massive galaxies to about 
20 per cent for galaxies with stellar masses 
${\rm \la 10^{9.5}\,M_{\odot}}$. 
These results imply that the evolution of star-forming galaxies 
is driven primarily by their intrinsic properties and is largely 
independent of their environment over a large range of local 
galaxy density.

\end{abstract}


\section{Introduction}

One of the central problems in astronomy is that of galaxy formation
and evolution: when were the various parts of galaxies assembled, 
when were the stars formed, and how did these processes depend on
environment? At low redshift the dependence of many galaxy properties
on environment is well established.  Dense environments contain a
larger fraction of red, passive galaxies, while low density
environments contain more blue, star-forming galaxies (e.g.,
\citealt{Dressler80,poggianti99,lewis02,gomez03}). The distributions
of colour and H$\alpha$ equivalent width, and hence inferred
star-formation histories, are found to be bimodal
\citep{baldry04,balogh04}.  Surprisingly, the two sequences vary 
little with environment, except in terms of their relative proportions
\citep{balogh04,balogh04bimodal,baldry06}.  However, some studies find
that star-forming galaxies in clusters tend to have a reduced global
star formation rate with respect to field galaxies of the same
morphological type \citep{KoKe04}.

The chemical abundances of stars and interstellar gas within galaxies 
provide a fundamental tool for tracing the evolution of their stellar 
and metal content.  These abundances depend on various physical
processes, such as the star formation history, gas outflows and 
inflows, etc. Variation in the chemical enrichment of galaxies may
potentially provide insight into how environment affects galaxy
evolution.  Measurements of this variation can thus assist in
constraining the likely scenarios of galaxy evolution.

The correlation between galaxy metallicity and luminosity is one 
of the most significant observational results in chemical evolution
studies (e.g., \citealt{lequeux79}; \citealt*{SKH89,ZKH94};
\citealt{RM95,MS02,lee06}).  The analysis of large samples of
star-forming galaxies shows that metallicity correlates with
luminosity over $\sim10$ magnitudes, with a factor of $\sim100$ 
increase in metallicity \citep{lamareille04}. 
The metallicity relation is tighter versus stellar mass than 
luminosity \citep{tremonti04}.

Spectrophotometric observations of {H{\sc ii}} regions for 
a sample of nine Virgo cluster spiral galaxies have shown 
that {H{\sc i}}-deficient objects, near the cluster core, 
appear to have higher oxygen abundance than field galaxies 
of comparable luminosity or morphology (\citealt{skillman96}). 
Oxygen abundances of spirals at the periphery of the cluster 
with normal {H{\sc i}} properties are however indistinguishable 
from those of field galaxies (\citealt*{SSK91}; see also 
\citealt{DC06} for similar results). 
Due to their lower gravitational potentials, dwarf galaxies 
ought to be more sensitive to their surroundings and should 
especially be less able to retain their gaseous contents. 
For a sample of dwarf star forming galaxies located at 
different environments, \citet{Vilchez95} found that galaxies 
located in low density regions tend to display higher 
excitation-sensitive emission line ratios, with higher 
emission line equivalent widths and total luminosities. 
Dwarf galaxies appear however to follow the same 
metallicity-luminosity relation, i.e., at a given luminosity, 
there is no systematic difference in oxygen abundance between 
cluster and field star-forming dwarf galaxies. 
\citet*{LMR03} have confirmed this result for a sample of 
dwarf irregulars in the Virgo cluster. A subsample of Virgo 
star-forming galaxies display a much lower baryonic gas 
fraction than their counterparts in the field at comparable 
oxygen abundances. They argue that the observed gas-poor 
star forming galaxies have being stripped of their gaseous 
content without significant effect on their luminosities and 
metallicities.

The large dispersion in the chemical properties of field galaxies 
(see e.g., \citealt{ZKH94}), and previous small sample sizes in 
dense regions make it difficult to draw definitive conclusions 
regarding the effect of environment on the chemical properties 
of galaxies. The large data sets provided by the Sloan Digital 
Sky Survey (SDSS) now allow the environmental effects on galaxy 
properties to be followed statistically over the full range of 
environments, from the sparse field to dense cluster cores, at 
least for bright galaxies. Additionally, the homogeneity of the 
SDSS sample enables us to sample the entire range of galactic 
environments, in a uniform manner. We can now, therefore, expand 
the study of galaxy chemical properties from the usual cluster 
versus field comparison to general environment, in this case 
characterised by local density, irrespective of actual cluster 
or group membership. This will help strongly in constraining
the physical processes responsible for determining galaxy 
properties. In this paper, we analyze the variation in oxygen 
abundance of the interstellar star-forming gas for a large 
sample of local galaxies as a function of their stellar mass 
and environment, in order to investigate environmental effects 
on the chemical content of galaxies, and how it depends on
galaxy mass.

The paper is organized as follows. In Section~\ref{data} 
we describe the sample selection. In Section~\ref{results}, 
we investigate the environmental dependence of the stellar 
mass versus gas-phase oxygen abundance relationship.
The implications of our results, and our conclusions are
summarized in Section~\ref{summary}. The cosmological model 
with ${\rm H_0=70\,km\,s^{-1}}$, $\Omega_{m}=0.3$, 
$\Omega_{\Lambda}=0.7$  has been adopted throughout the paper.

\section{Data and sample selection}
\label{data}

The SDSS is a project with a dedicated 2.5-m telescope 
designed to image $10^4$ sq.\ deg.\ and take spectra of 
$10^6$ objects \citep{york00}.  The imaging covers five 
broadbands, $ugriz$, with effective wavelengths from 
350\,nm to 900\,nm. The spectra are obtained using a 
fibre-fed spectrograph with $3''$ apertures. 
This aperture size corresponds to 3\,kpc at $z\sim0.05$.
The wavelength range is from 380 to 920\,nm and the 
resolution is $\lambda/\Delta\lambda\sim1800$.

The majority of spectra are taken of galaxies with $r<17.77$, 
called the main galaxy sample (MGS) \citep{strauss02}. 
The sample analysed here was selected from SDSS Data Release 
Four \citep{adelman-mccarthy06}.
{\bf
The initial sample consisted of 151\,168 MGS galaxies in the 
redshift range 0.010--0.085. The sample is volume-limited for 
galaxies brighter than $M_r=-20$, which is the density defining 
population. For magnitudes fainter than this limit, the sample 
is magnitude-limited.
}
Data on these galaxies were taken from the standard {\small PHOTO} 
and spectroscopic data reduction pipelines \citep{stoughton02}, 
along with additional measurements from the MPA Garching DR4 data 
\citep{kauffmann03A,brinchmann04,tremonti04} and environmental 
measurements \citep{baldry06}. 

Environmental density measurements were determined using an 
$N$th nearest neighbour technique. Galaxies more luminous 
than $M_r<-20$ were used for the density defining population 
(DDP): with an average space density of $0.005{\rm\,Mpc}^{-3}$. 
The environmental density for each galaxy is then given by
\begin{equation}
\log\Sigma = 
       \frac{1}{2} \log \left(\frac{4}{\pi d_4^2}\right) + 
       \frac{1}{2} \log \left(\frac{5}{\pi d_5^2}\right)
\label{eqn:sigma_n}
\end{equation}
where $d_4$ and $d_5$ are the projected distances to the 4th 
and 5th nearest DDP neighbours within $1000{\rm\,km\,s}^{-1}$.  
For typical galaxies, $\Sigma$ ranges from $0.05{\rm\,Mpc}^{-2}$, 
typical of void-like regions, to $20{\rm\,Mpc}^{-2}$, typical 
of the centres of galaxy clusters. A best-estimate $\Sigma$, 
and minimum and maximum values, were determined taking account 
of edge effects and galaxies that had not been observed 
spectroscopically. 
For example, $\Sigma_{\rm max}$ was determined by using the 
smaller of (i) the distance to the nearest edge and (ii) the 
distance to the 4th/5th nearest photometrically confirmed 
neighbours. See \citet{baldry06} for details. 


From this sample, only galaxies for which emission lines are 
well detected were retained. We require galaxies included in 
our star-forming sample to have lines of {\oii}, {\oiiis}, 
{\hbeta}, {\halpha}, and {\nii} detected at greater than 
$5\sigma$, reducing the sample to 68\,603. Galaxies with 
$z<0.03$ are excluded from the sample due to the blue 
wavelength cutoff of the spectrograph. We have distinguished 
star-forming galaxies from non-thermal sources, such as active 
galactic nuclei (AGN), using the classical diagnostic ratios 
of two pairs of relatively strong emission lines, i.e., 
{\oiiis}/{\hbeta} vs. {\nii}/{\halpha} diagrams 
\citep*{BPT81,VO87}. We have used the empirical demarcation 
line, separating star-forming galaxies from AGN, provided by 
\citet{kauffmann03agn}. The selection of star-forming galaxies 
reduced the sample to 51\,934 galaxies. In order to avoid edge 
effects, galaxies with 
$(\log \Sigma_{\rm max} - \log \Sigma) > 0.4$ were excluded 
from the analysis. This reduced the sample to 45\,107 galaxies.

For fibre-fed spectroscopy, the fraction of the light of a 
galaxy captured within the fiber will depend on its redshift, 
intrinsic size, and surface brightness profile, as well as the 
size of the fibre and seeing during the observation. 
\citet{kewley05} have shown that using the spectra of the inner 
parts of galaxies could have a substantial systematic effects 
on the estimate of their global properties. To reduce systematic 
and random errors from aperture effects, they recommended 
selecting SDSS galaxies with $z>0.04$. Doing so, reduces the 
size of the sample to 37\,866 galaxies.

Gas-phase oxygen abundances are taken from the catalogues of derived
physical properties for DR4 galaxies released by the MPA Garching
group \citep{tremonti04}. Briefly, gas-phase oxygen abundances are
estimated based on simultaneous fits of all the most prominent
emission line fluxes ({\oii}, {\hbeta}, {\oiiis}, {\halpha}, {\nii},
{\sii}) with a model designed for the interpretation of integrated
galaxy spectra based on a combination of population synthesis and
photoionization codes \citep{CL01}. The likelihood distribution of 
the metallicity of each galaxy in the sample is calculated, based 
on comparisons with a large library of models corresponding to 
different assumptions about the effective interstellar star-forming 
gas parameters. 
{\bf The median of this distribution is adopted as the best estimate 
of the galaxy metallicity. The estimated metallicities show good 
agreement with those estimated using the so-called strong emission 
line method. The width of the likelihood distribution provides 
a measure of the error. The median $1\sigma$ error for our final 
sample of star-forming galaxies is $\sim 0.04$ dex.}

{\bf The computation of stellar masses by the MPA Garching group 
is obtained by fitting a grid of population synthesis models, 
including bursts, to the spectral features D4000 and H$\delta$ 
absorption. The predicted colours are then compared with broad-band 
photometry to estimate dust attenuation. Stellar mass-to-light 
ratios are determined and applied to the Petrosian $z$-band 
magnitude. The nominal $1\sigma$ random errors are typically 
$<0.1$ dex. For details see \citet{kauffmann03A}. Assuming most 
galaxies have stellar mass estimates with relative uncertainties 
less than 0.2 dex, these have little impact on our results.}

\begin{figure*}
\includegraphics[clip=,width=0.49\textwidth]{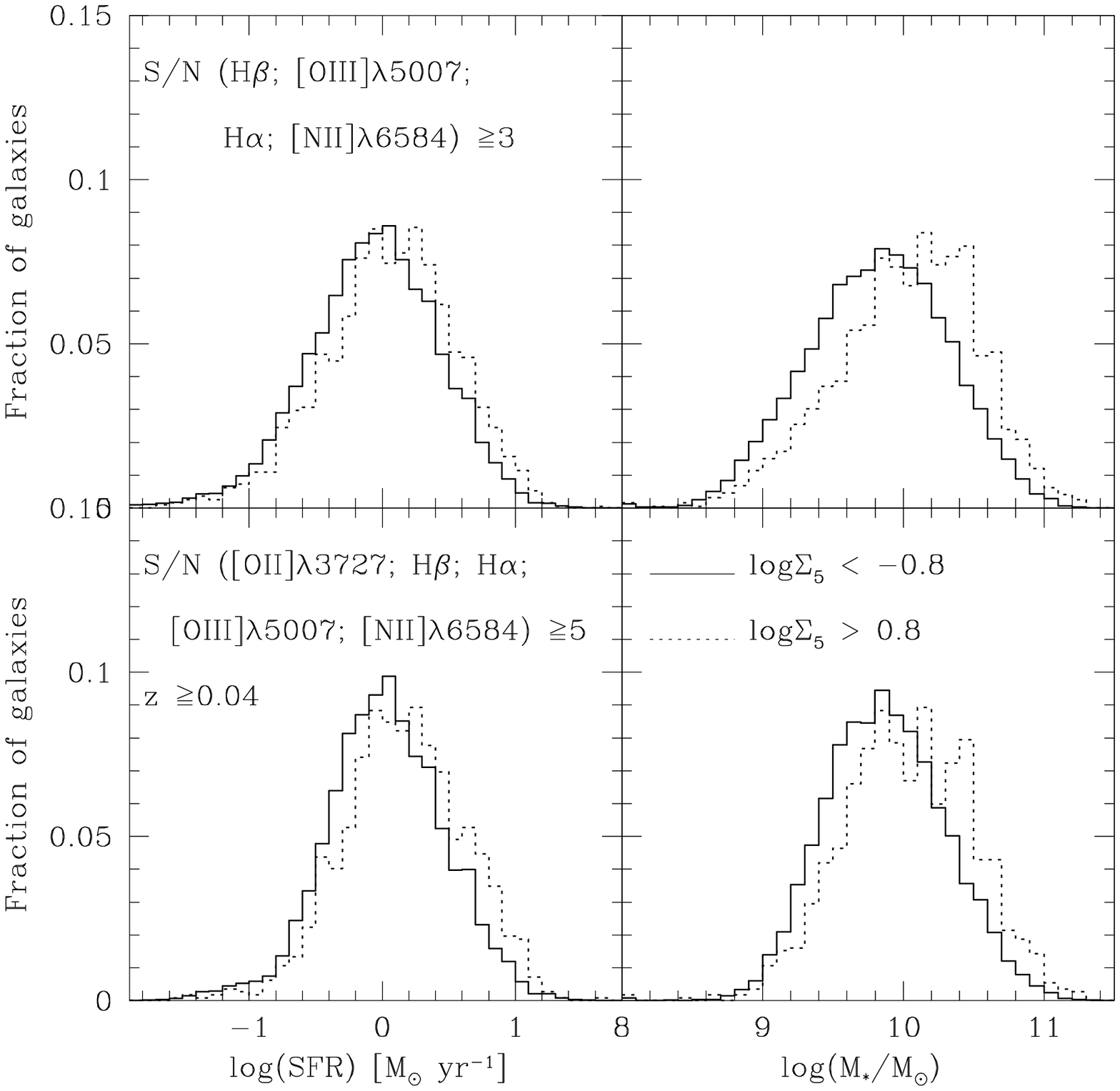}
\includegraphics[clip=,width=0.49\textwidth]{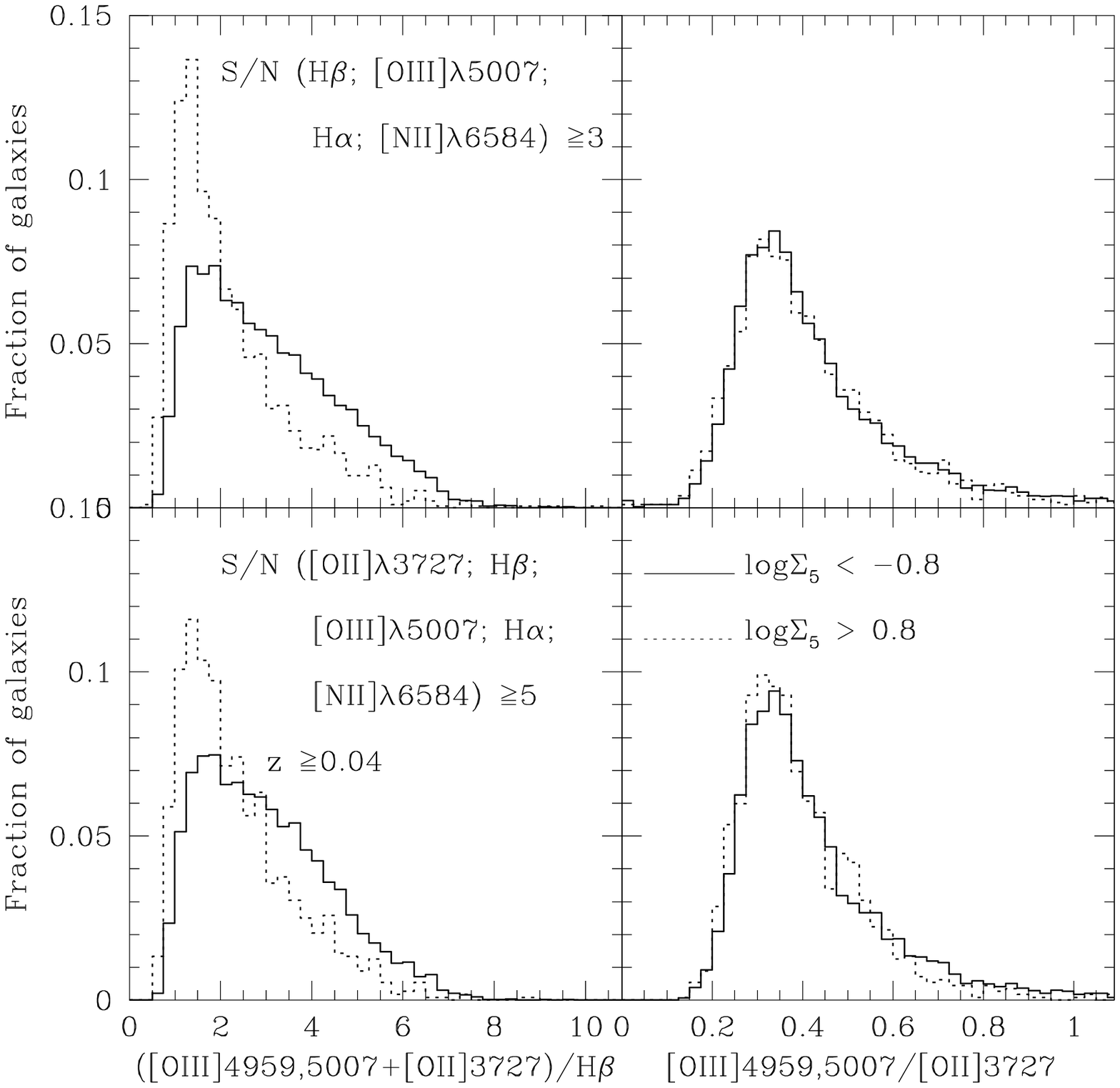}
\caption{ Left: distributions of star formation rate and stellar 
mass for the final selected sample (the lower panels), and a 
comparison sample of star-forming galaxies selected to have the 
emission lines needed to classify the ionizing source, i.e.,
{\hbeta}, {\oiiis}, {\halpha}, and {\nii}, detected with {\snr} 
greater than 3 (the upper panels). Solid lines show the 
distribution of the properties of galaxies located in void-like 
regions, i.e., $\log \Sigma < -0.8$, while dotted lines show the 
distribution of the properties of those situated in dense region,
$\log \Sigma > 0.8$. Right: same as in the right panel except 
for the abundance-sensitive ({\oiii}+{\oii})/{\hbeta} ratio, 
and excitation-sensitive diagnostic ratio {\oiii}/{\oii}.}
\label{selection}
\end{figure*}

{\bf
Selecting only galaxies with high {\snr} emission lines could 
potentially bias the distribution of galaxy properties in the 
selected final sample in such a way to affect the intrinsic 
correlations between galaxy properties that we aim to investigate. 
Therefore it is important to assess to what extent the final 
sample of star-forming galaxies covers similar regions in the 
parameter space as star-forming galaxies in original sample. 
To ensure that the final sample of star-forming galaxies is 
representative of the parent sample, we compare the distribution 
of galaxy properties in the final sample with those of a more
lenient selection. These were selected from the initial MGS 
sample, as those with {\snr} larger than 3 in emission lines 
needed to classify the ionizing source, i.e., {\oiiis}, 
{\hbeta}, {\halpha}, and {\nii}.

Fig.~\ref{selection} shows the distributions of stellar mass, 
star formation rate, abundance-sensitive diagnostic ratio 
$({\oiii}+{\oii})/{\hbeta}$, and excitation-sensitive diagnostic 
ratio ${\oiii}/{\oii}$ for both the final and the parent samples 
of star-forming galaxies. The distributions of the properties of 
the star-forming galaxies in the parent sample are shown in the 
upper panels, and those of the final sample are shown in the 
bottom panels. Solid lines show the distribution of the 
properties of galaxies situated in rarefied field regions, i.e., 
$\log \Sigma < -0.8$, and dotted lines show the distribution 
of the properties of those situated in dense region, i.e., 
$\log \Sigma > 0.8$. The figure shows clearly that the final 
sample has galaxy parameter distributions similar to those of 
the parent star-forming galaxy sample. Most importantly, the 
differences with respect to the environment are the same for 
the two different selections.
Observed abundance- and excitation-sensitive diagnostic ratios 
are distributed similarly in both samples. 
However the redshift cut, imposed to minimize the effects of the 
aperture bias, appears to slightly affect the extent of the low 
end tails of stellar mass and star formation rate distributions,
i.e., the low ends of both distributions for the final sample 
are less extended than for the parent star-forming galaxy sample. 
The cut on redshift excludes low stellar mass galaxies, as those 
galaxies tend to be at low redshift end.  
Those galaxies represent however a small fraction of the sample. 
The high ends of the star formation rate and the stellar mass 
distributions are however unaffected by the redshift cut. 
The similar distributions of both star-forming galaxy samples 
suggest that our final sample is a fair representation of the 
parent sample in terms of its stellar mass, star formation 
activity, interstellar gas-phase properties. The correlations 
we aim to investigate are therefore expected to not be severely 
biased by the selection procedure of the final sample of 
star-forming galaxies. 
}

\begin{figure}
\includegraphics[clip=,width=0.5\textwidth]{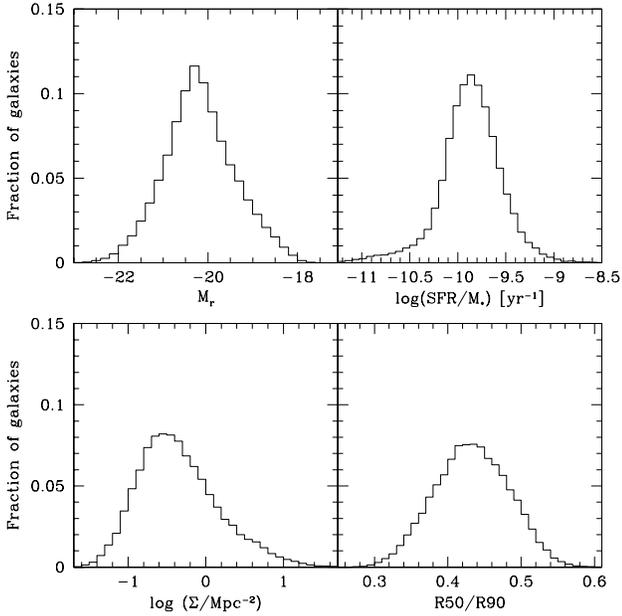}
\caption{Distribution of general properties of our sample of 
star-forming galaxies. The upper panels show the absolute $r$-band 
magnitude, and specific star formation rate. The lower panels show 
the distributions of local galaxy density, and (inverse) 
concentration index.}
\label{prop_dist}
\end{figure}

\begin{figure*}
\includegraphics[clip=,angle=90,width=0.90\textwidth,height=12.cm]
{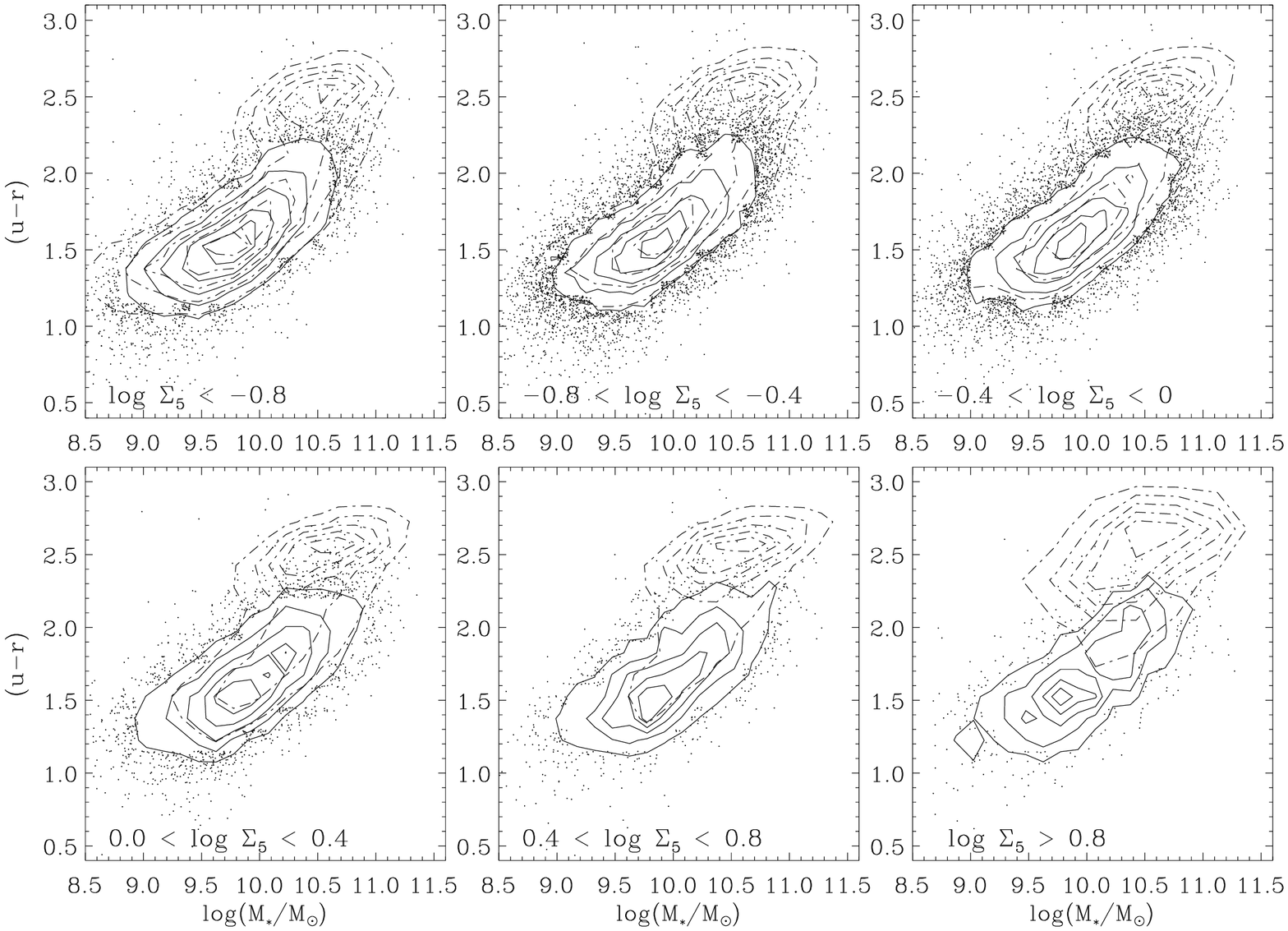}
\caption{Colour vs. stellar mass relations for different environments. 
Solid contours and points show the distribution of the selected 
galaxies discussed in the remainder of the paper, and the 
dashed-dotted lines represent the initial complete sample selected 
from SDSS DR4. It is clear that we are selecting galaxies that 
belong to the blue sequence.}
\label{cmr_sel}
\end{figure*}

{\bf
The distributions of global properties for our final sample, i.e., 
$r$-band absolute magnitudes, specific star formation rate, local 
galaxy density, and the concentration index, defined as the ratio 
of the radii enclosing 50\% ($R_{50}$) and 90\% ($R_{90}$) of the 
Petrosian $r$-band galaxy light, are shown in Fig.~\ref{prop_dist}.
}

Fig. \ref{cmr_sel} shows the relationship between the integrated 
$(u-r)$ colour and the stellar mass for six local galaxy density 
bins ranging from $\log\Sigma < -0.8$ to $\log\Sigma > 0.8$
(see below). The distribution of the final sample of galaxies 
is shown as solid contours and points. The full parent sample 
of galaxies in the redshift range 0.010--0.085 is shown as the 
dashed-dotted contours. The bimodal nature of the colour-stellar 
mass relation for the complete sample is clear: red galaxies 
dominate in dense regions and at higher masses, while the 
dominance of blue galaxies is clear at lower masses and for 
lower local densities (see \citep{baldry06} for more details). 
The selected galaxies are distributed along the blue sequence. 
This is no surprise, as star-forming galaxies with well 
detected emission-lines tend to have blue colours.

\section{Results}
\label{results}

\begin{figure*}
\includegraphics[clip=,width=0.92\textwidth,height=14.cm]
{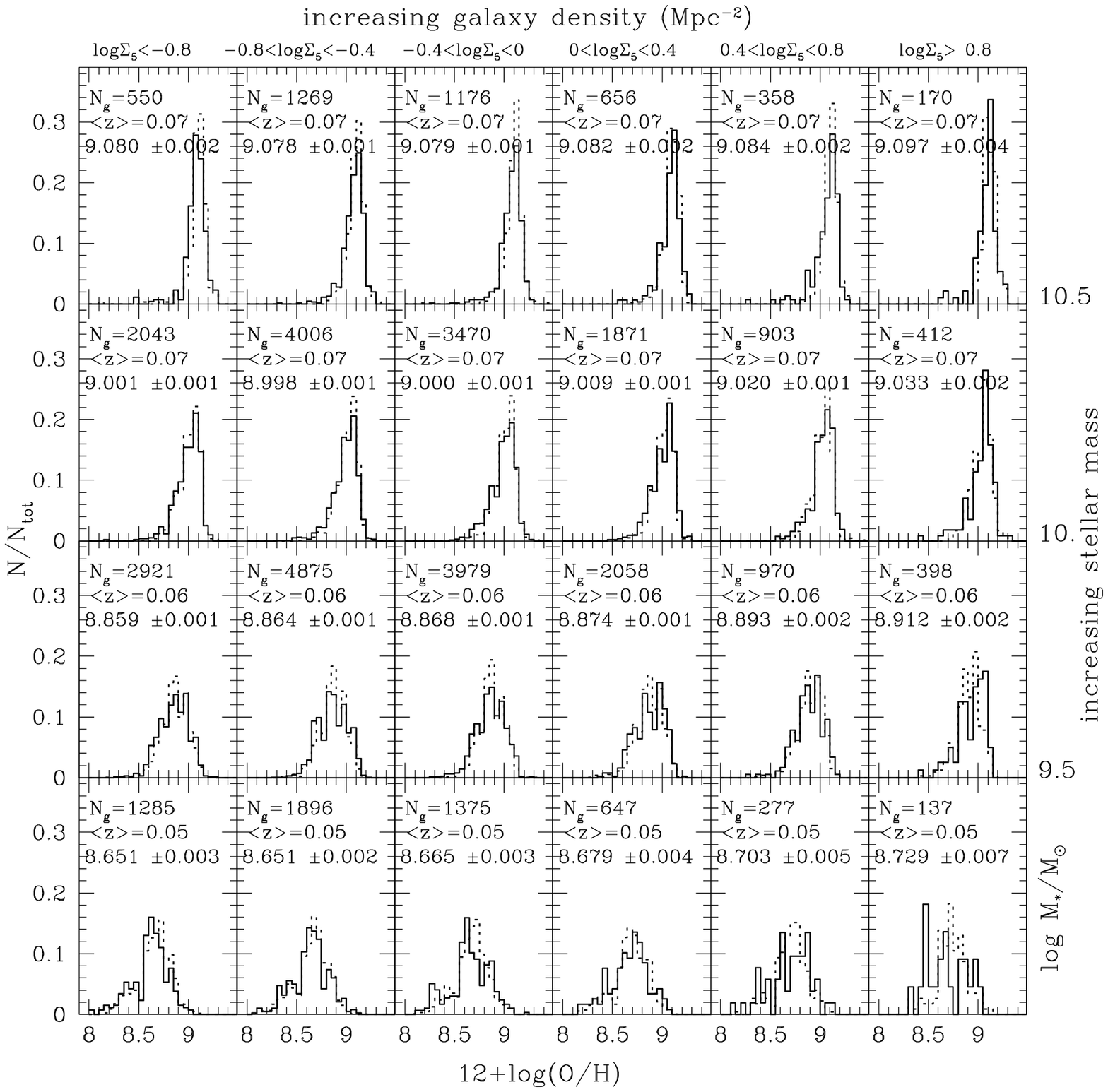}
\caption{
Distributions of galaxy gas-phase oxygen abundance for the 
indicated range of stellar mass (right axis) and the range 
of local galaxy density in units of Mpc$^{-2}$ (top axis).  
The solid lines represent galaxies with concentration indices 
lower than 0.4 (nominally early types), while the dotted 
lines represent those larger than 0.4 (nominally late types).
The number of galaxies in the bin, the mean of their redshift
distribution, and average metallicity, are given in each 
panel.}
\label{ohdist}
\end{figure*}

The selected sample of star-forming galaxies was divided 
into bins of local galaxy density and stellar mass in 
order to determine the variation of gas-phase oxygen 
abundances as a function of these quantities.
Figure~\ref{ohdist} shows the gas-phase oxygen abundance 
distribution of galaxies in bins of galaxy local density 
and stellar mass. The median local galaxy density in the 
selected sample is ${\rm \Sigma \sim 0.4\,Mpc^{-2}}$; thus, 
the lowest local galaxy density bin, with a median density 
of ${\rm \Sigma \sim 0.1\,Mpc^{-2}}$ is underdense by a 
factor of $\sim 4$, while the densest bin with a median 
density of ${\rm \Sigma \sim 10\,Mpc^{-2}}$, corresponds 
to the typical galaxy density found in the inner regions 
of galaxy clusters. 

Our selection function preferentially selects morphologically
late-type galaxies. However because the strongest effect with 
environment is the early/late-type galaxy fraction; we split 
the sample into nominally early and late-type galaxies. 
The (inverse) concentration index, is used as a proxy for 
galaxy morphology: with $R_{50}/R_{90}\sim 0.4$ marking the 
boundary between early and late types 
\citep{shimasaku01,strateva01}. \citet{baldry06} have found 
that a concentration index around $\sim 0.4$ is a natural 
dividing line between red and blue galaxy populations for 
all stellar masses. \citet{driver06} also show that galaxies 
naturally divide into two populations in the 
concentration -- colour plane, which can be associated with 
late- and early-types. Solid (dotted) line histograms in 
Fig.~\ref{ohdist} represent gas-phase oxygen abundance 
distributions for galaxies with concentration index lower 
(higher) than 0.4. The distributions of gas-phase oxygen 
abundances of star-forming galaxies with concentration 
indices larger (late-type population) and lower (early-type 
population) than the dividing value are indistinguishable 
for all environments.

In all local galaxy density bins, the median of the gas-phase 
oxygen abundance distribution increases with increasing stellar 
mass, i.e., the so-called stellar mass -- metallicity relation 
is present in all environments. The dispersion of the gas-phase 
oxygen abundance distribution increases at lower stellar mass, 
where the fraction of star-forming galaxies increases 
(e.g.\ \citealt{balogh04}), but appears to be insensitive to 
environment. 

Most remarkable is the mass dependence of the environmental 
effect on galaxy gas-phase oxygen abundance. 
For massive galaxies, i.e., stellar masses larger than 
${\rm \sim 10^{10.5}M_{\odot}}$, the oxygen abundance 
distribution median does not show any significant dependence
on the environment. However, for low mass galaxies 
(${\rm \sim 10^{9.5}M_{\odot}}$), the abundance distribution 
median increases by 0.06--0.08 dex from low density 
environments, i.e., $\log \Sigma \la -0.8$, to highly dense 
regions, i.e., $\log \Sigma \ga 0.8$.

\begin{figure*}
\includegraphics[clip=,angle=90,width=0.90\textwidth]
{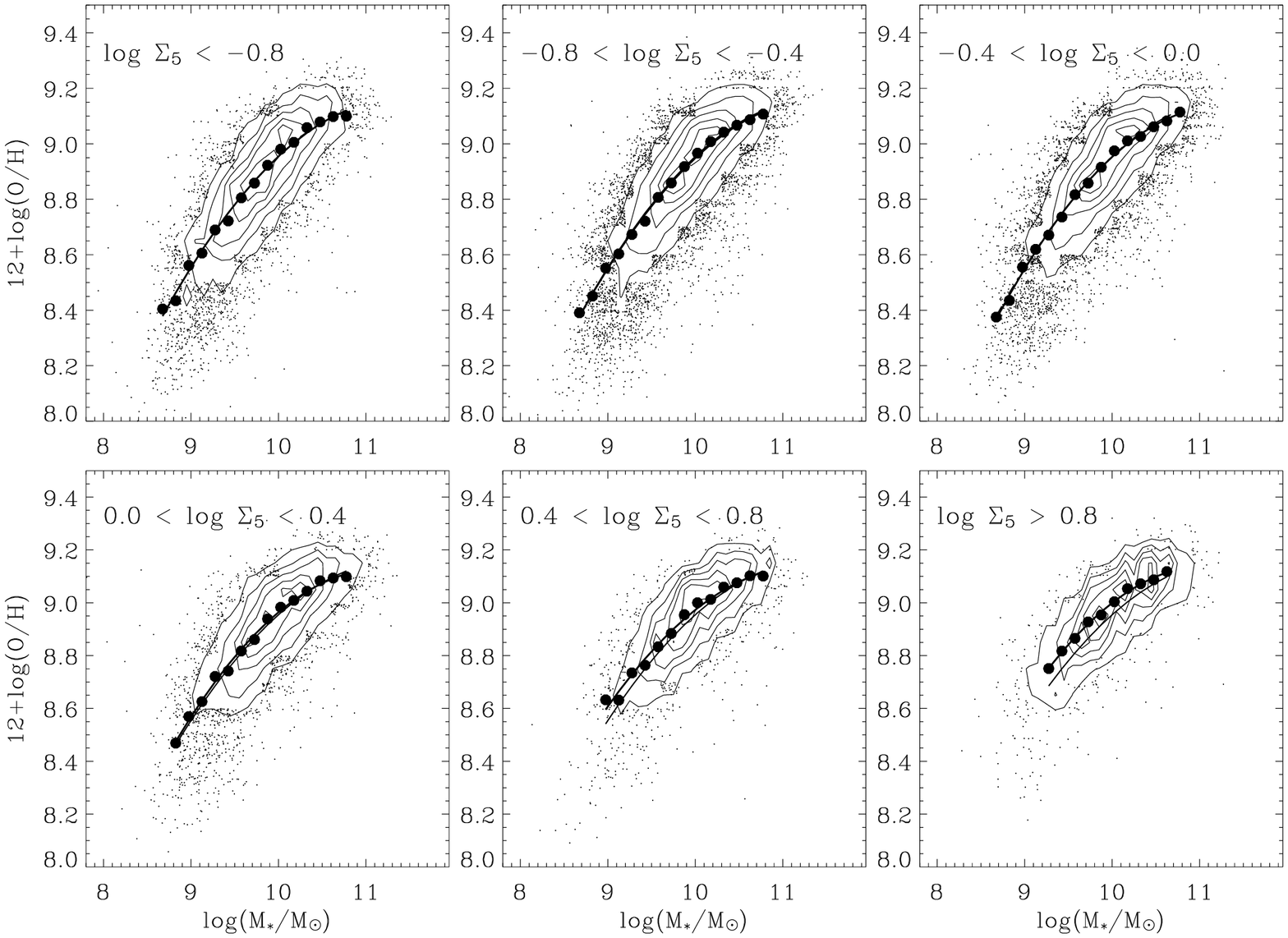}
\caption{Gas-phase oxygen abundance versus stellar mass 
relation for different local galaxy density bins ranging
from void-like environments (upper left panel) to 
cluster-like environments (bottom right panel). 
{\bf The median $1\sigma$ metallicity error for the galaxy sample 
is $\sim 0.04$ dex.} The large filled points represent oxygen 
abundance medians in bins of 0.15 dex in mass that include 
at least 100 data points. In each panel, the thick line 
shows a polynomial fit to the data, and the thin line shows 
the fit to the stellar mass vs. metallicity for galaxies 
with $\log\Sigma < -0.8$.}
\label{ms_oh_env}
\end{figure*}

To determine the variation of the stellar mass versus gas-phase 
oxygen abundance relation as a function of the environment, we 
have divided the star-forming galaxy sample into local galaxy 
density bins. Figure~\ref{ms_oh_env} shows the relationship 
for six local density bins ranging from $\log\Sigma < -0.8$ 
to $\log\Sigma > 0.8$. In each panel, the large filled circles 
show the median gas-phase oxygen abundance in bins of 0.15 dex 
in stellar mass containing more than $50$ galaxies, and the 
thick line shows a polynomial fit to these points. 
The thin solid line shows the fit to the stellar mass versus 
oxygen abundance for galaxies in low density regions, i.e., 
$\log\Sigma < -0.8$ (only visible in the higher density bins). 
The correlation is relatively steep from 
$10^{8.5} {\rm M_{\odot}}$ to $10^{10.5} {\rm M_{\odot}}$, 
but flattens for more massive galaxies (as discussed by 
\citealt{tremonti04}). 

\begin{figure*}
\includegraphics[clip=,angle=90,width=0.75\textwidth]
{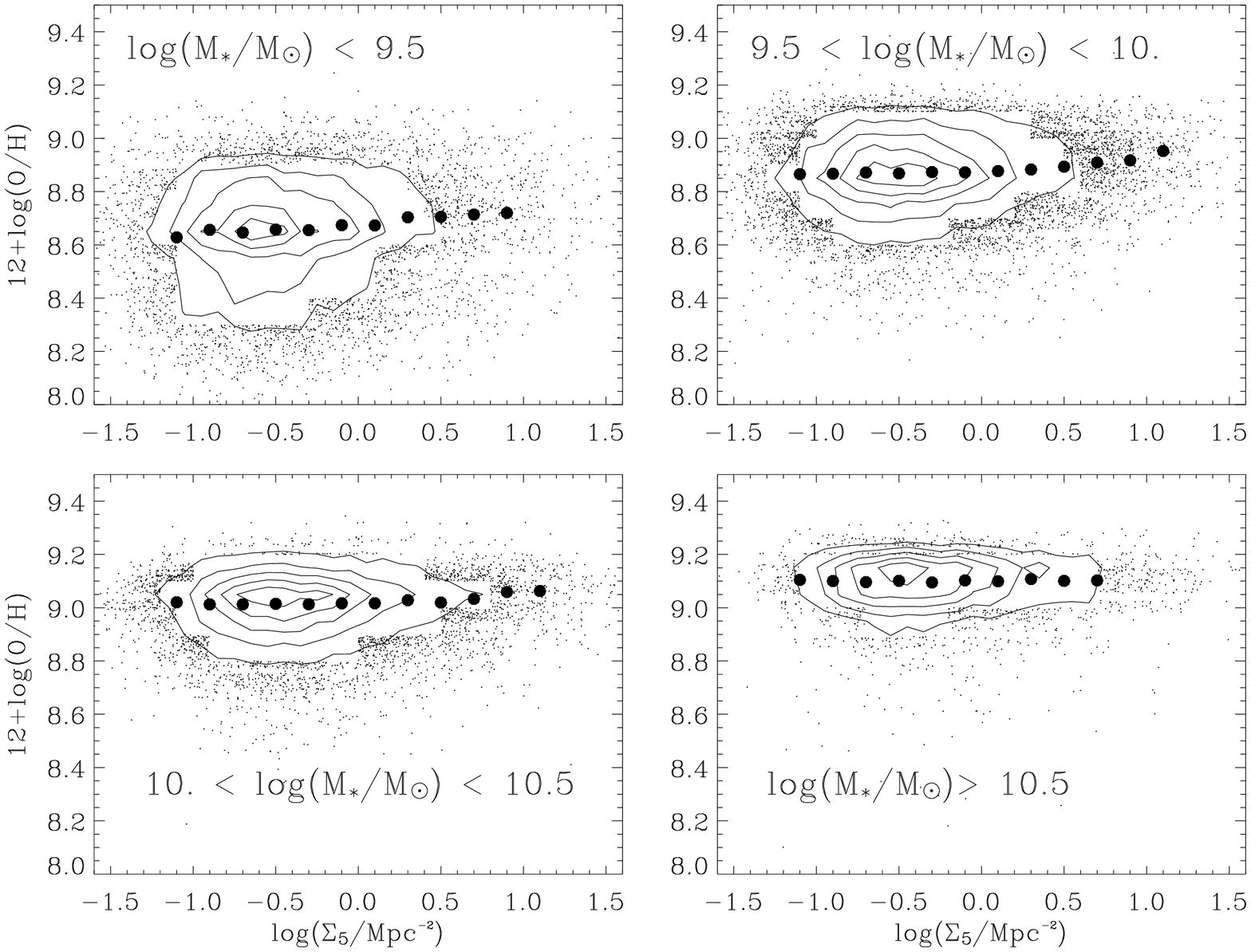}
\caption{Gas-phase oxygen abundance versus projected local 
galaxy density for different stellar mass bins. 
Large filled circles show the median in bins of 0.15 dex 
in projected local galaxy density.}
\label{oh_density}
\end{figure*}

Most striking of all is the similarity of the correlation 
between stellar mass and gas-phase oxygen abundance over 
a large range of local galaxy densities, from 
$\log\Sigma < -0.8$, typical of void-like environments, to 
$\log\Sigma \approx 0.8$, typical of the outer regions of 
galaxy clusters. This is in agreement with previous findings, 
that star-forming galaxies in the periphery of the Virgo 
cluster exhibit similar oxygen abundances to those of field 
galaxies with similar luminosities \citep{skillman96,pilyugin02}.
The spread of oxygen abundances about the median is identical 
at all environments across the entire galaxy stellar mass 
range. The bottom right panel of Fig.~\ref{ms_oh_env} shows, 
however, that stellar mass versus oxygen abundance relationship 
at denser environments, i.e., $\log\Sigma > 0.8$ typical of 
the inner parts of galaxy clusters, is shallower than the 
relationship for galaxies in less dense regions. 
Gas-phase oxygen abundances show a mass-dependent shift to 
higher values that is larger for low stellar mass galaxies. 
This seems to disagree with the results of \citet{skillman96} 
who found that the interstellar star-forming gas of bright
galaxies near the core of the Virgo galaxy cluster appear 
to be overabundant by 0.2--0.3 dex in comparison with field 
galaxies with similar luminosities. \citet{pilyugin02} have 
confirmed the higher abundances for the same Virgo galaxies 
studied by \citeauthor{skillman96}, although they have found 
counterparts for both the periphery and core cluster objects 
among field star-forming galaxies. They have concluded that
if there is a difference in the abundance properties of the 
Virgo and field bright spirals, this difference appears to 
be small, in agreement with our results.

Figure~\ref{oh_density} shows the variation of gas-phase 
oxygen abundance as a function of local galaxy density for 
different stellar mass bins. Over a factor $\sim 100$ in 
local galaxy density the median gas-phase oxygen abundance 
changes by only $0.02-0.08$ dex, depending on stellar mass, 
being greater at the low stellar mass end. These trends are 
weak, relative to the stellar mass dependence: the median 
gas-phase oxygen abundance is $\sim 0.7$ dex higher in the 
massive galaxies, compared with the low mass galaxies in 
the same environment.

\begin{figure}
\includegraphics[clip=,angle=90,width=0.5\textwidth]
{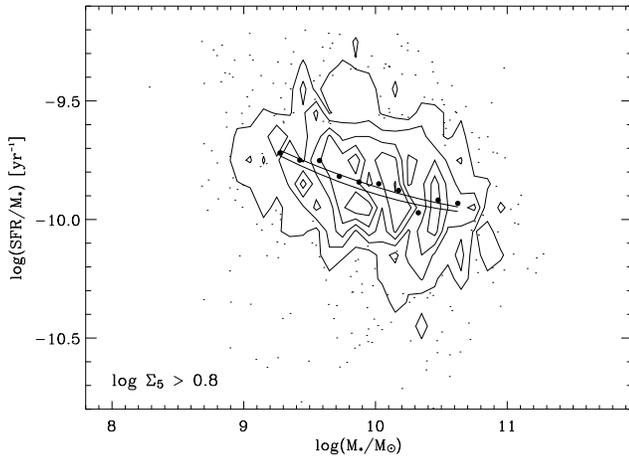}
\caption{Specific star formation rate versus stellar mass 
for galaxies in dense regions, i.e., 
$\log\Sigma > 0.8$. The large filled points 
represent the median in bins of 0.1 dex in mass. The thick 
line shows a polynomial fit to the data, and the thin line 
shows the fit to the same relation for galaxies with 
$\log\Sigma < -0.8$.}
\label{nsfr_density}
\end{figure}

Figure~\ref{nsfr_density} shows the relationship between the 
specific star formation rate and the stellar mass for galaxies 
in dense environments, $\log\Sigma > 0.8$. We use the star 
formation rates corrected to total computed by 
\citet{brinchmann04}. Note that this sample only includes 
galaxies which are actively star forming, due to our 
requirement on significant emission-line detections. However, 
it still meaningful to consider star-forming galaxies alone 
in this diagram, due to the bimodality in specific star 
formation rates, e.g., as seen by \citet{balogh04} considering 
the H$\alpha$ equivalent width distribution. 
As in Fig.~\ref{ms_oh_env}, the thick line shows a polynomial 
fit to the variation of the specific star formation rate median, 
shown as large field circles, as a function of the stellar mass. 
The thin solid line shows the fit to the same correlation for 
galaxies situated in low density regions, i.e., 
$\log\Sigma < -0.8$.

Similar to the stellar mass versus oxygen abundance relation, 
the correlation between the specific star formation rate and 
the stellar mass of star-forming galaxies shows very little 
dependence on local galaxy density from void-like environments, 
i.e., $\log\Sigma <- 0.8$, to the periphery of galaxy clusters, 
i.e., $\log\Sigma \approx 0.8$. For star-forming galaxies in 
dense regions, i.e., $\log\Sigma > 0.8$, the median of the 
specific star formation rate distribution is, however, 
$\sim 0.03$ dex higher than for galaxies with similar stellar 
masses in less dense regions.

\section{Summary \& Discussions}
\label{summary}

Using a complete sample of 151\,168 galaxies in the redshift 
range of $0.01<z<0.085$ from the SDSS DR4, we draw a sample 
of 37\,866 star-forming galaxies in the the redshift range 
of $0.04 < z < 0.085$ to examine how oxygen abundances of 
the interstellar star-forming gas relate to environment. 
We use the projected distance to the fourth and fifth nearest 
neighbours as an indicator of local galaxy density. The stellar 
mass versus gas-phase oxygen abundance relationship appears 
to be environment-free over a large range of local galaxy 
densities. For galaxies located in environments with local 
galaxy densities typical of the inner regions of galaxy 
clusters, the interstellar star-forming gas is found to be 
slightly overabundant compared to what is found for their 
counterparts of similar stellar masses in less dense environments. 
This change in oxygen abundance increases as galaxy stellar mass 
decreases, ranging from $\sim 0.2$ dex for galaxies with stellar 
mass larger than ${\rm \sim 10^{10}\,M_{\odot}}$ to 
$\sim 0.08$ dex for galaxies with stellar mass around 
${\rm \sim 10^{9}\,M_{\odot}}$.

Over that last few years, observational evidence has been 
accumulating that the dominant evolution in the galaxy 
population properties as a function of environment is the 
change of the ratio between star-forming/blue galaxies 
and passive/red galaxies (e.g.\ \citealt{Dressler80}; 
\citealt{baldry06} and reference therein). The environmental 
dependence of different galaxy properties, i.e., colour, 
concentration index, colour gradient, are found to be almost 
entirely due to the dependences of galaxy morphology and 
luminosity on the environment: when morphology and luminosity 
are fixed, galaxy physical properties are nearly independent 
of local galaxy density \citep{park06}. The scaling relations 
for both passive and star-forming galaxies are virtually 
unaffected by environment. Studies of the fundamental plane 
have shown that the properties of early-type galaxies are 
nearly independent of environment (e.g., 
\citealt{dressler87,bernardi03C}). The scaling relations for 
star-forming galaxies are virtually unchanged with environment, 
whilst over the same density range the fraction of star-forming 
galaxies is changing strongly 
(e.g.\ \citealt{balogh04,blanton05enviro,baldry06}).

Our finding that oxygen abundances of galaxies with a given 
stellar mass depend weakly on environment is consistent with 
(i) the weak dependence on environment of the mean colour, at 
a given luminosity, of blue galaxies that are still actively 
growing and evolving, compared with the dependence on 
luminosity, the strongest effect also being seen for faint 
galaxies \citep{balogh04bimodal}, and (ii) the lack of any 
environmental effect on the distribution of the emission line 
equivalent width for star-forming galaxies \citep{balogh04}.

The weak dependence of the star-forming gas oxygen abundance 
shown here is a further suggestion that the primary physical 
driver(s) of galaxy evolution must depend primarily on 
galaxy intrinsic properties rather than on their environment.  
In addition, whatever mechanism is responsible for the 
increasing fraction of passive galaxies in dense environments, 
it must truncate their star formation on a short timescale to 
avoid affecting the relations of oxygen abundance and specific 
star-formation versus stellar mass for actively star-forming 
galaxies. 
Yet \citet{kauffmann04} have argued, based on the absence of 
dependence on environment of the correlations between different 
spectral indicators that probe star formation history on 
different timescales (${\rm SFR/M_{*}, D_{4000}, H\delta}$), 
that the decrease in star formation activity in dense 
environments occurs over long (${\rm \ga 1 Gyr}$) timescales. 
It is not clear however how slow quenching could affect the 
chemical properties of galaxies, and for which kind of star 
formation histories galaxies would move along the observed 
relations in different environments. Self-consistent modelling 
is clearly needed in order to set tight constraints on quenching 
timescales from observed chemical and spectro-photometric 
properties of galaxies. This is obviously well beyond the scope 
of this paper, and we will leave it for future investigations.

The modest change of oxygen abundances of star-forming galaxies 
in regions with local galaxy densities typical of the inner 
regions of galaxy clusters indicates however that the chemical 
evolution of galaxies is moderately modulated by the environment. 
Furthermore, the mechanism responsible for this change must be 
more efficient in low stellar mass galaxies, as the change of 
oxygen abundances is larger for those galaxies.
The relationship between stellar mass and oxygen abundance 
can be understood either as a sequence in astration, i.e., 
more massive galaxies are able to convert a larger fraction 
of their gas into stars than low mass counterparts, or as a 
depletion sequence, i.e., the efficiency of galaxies to retain 
their gas increases with galaxy mass. Based on the correlation 
between effective yield and baryonic mass for local 
star-forming galaxies, \citet{tremonti04} have argued that the 
most straightforward interpretation of the correlation is the 
selective loss of metals from galaxies with shallow potential 
wells via galactic winds. This suggests that the change of 
0.02--0.08 dex in the mean gas-phase oxygen abundance of 
star-forming galaxies as a function of environment can be 
explained by a reduced effectiveness of galactic winds in 
removing metals from the potential wells of low mass halos in 
dense regions. There is consequently more gas left to be turned 
into stars. Massive galaxies can however retain all their gas 
whether they are in the field or in a cluster. In this scenario, 
therefore, the efficiency of star formation within low mass 
halos is higher in the presence of a confining intergalactic 
medium, because supernovae are less efficient at removing 
gas from a galaxy if there is hot material surrounding the 
galaxy. 

An alternative explanation of the observed mass-dependent 
change in oxygen abundances of galaxies as a function of 
environment might be that low stellar mass galaxies have 
assembled their stellar contents at a faster rate in 
denser regions. Whatever triggers star formation in galaxy 
clusters, it causes it to happen rapidly, hence increasing 
the efficiency of recycling during star formation. 
Environment-driven processes, i.e., tidal interactions, 
mergers, could enhance the star formation activity in low 
mass galaxies in cluster environment. The observed systematic 
difference of specific star formation rate between field and 
cluster galaxies may be the signature of such an effect.

\bsp

\label{lastpage}

\end{document}